\documentstyle{EuroPhys}
\input EuroMacr
\begin{document} 

\euro{}{}{}{}
\Date{}
\shorttitle{B. ECHEBARRIA \etal PHASE INSTABILITIES IN HEXAGONAL PATTERNS}

\title{\bf Phase instabilities in hexagonal patterns \footnote{Dedicated to
	Prof. J. Casas-V\'azquez on the occasion of his 60th birthday}} 
\author {B. Echebarria and C. P\'erez-Garc\'{\i}a}
\institute{Departamento de F\'{\i}sica y Matem\'{a}tica Aplicada, 
	     Facultad de Ciencias, \\  
     Universidad de Navarra, E-31080 Pamplona, Navarra, Spain.}

\rec{}{}

\pacs{
\Pacs{}{47.54.+r}{Pattern selection}
\Pacs{}{47.20.Ky}{Nonlinearity}
\Pacs{}{42.65.Sf}{Dynamics of nonlinear optical systems}
}
    
\maketitle

\begin{abstract}
The general form of the amplitude equations for a hexagonal pattern including 
spatial terms is discussed. At the lowest order we obtain the {\em phase equation} 
for such patterns. The general expression
of the diffusion coefficients is given and the contributions of the
new spatial terms are analysed in this paper. From these coefficients
the phase stability regions in a hexagonal pattern are determined. 
In the case of Benard-Marangoni instability our results agree qualitatively
with numerical simulations performed recently.
\end{abstract}


Several systems out of equilibrium exhibit hexagonal patterns. 
Historically the cellular patterns reported by Benard almost a
century ago were the first nonequilibrium system showing this 
planform \cite{ben}. More
recently, hexagonal patterns were obtained in front solidification
\cite{wol}, in Rayleigh-Benard convection with 
non-Boussinesquian effects \cite{cil-bod-ste}, in 
Faraday crispation \cite{fau-gol}, in a nonlinear Kerr
medium \cite{d'al}, in a liquid crystal valve device
\cite{pam}, in chemical Turing patterns \cite{dek-swin},
in ferrofluids \cite{gai-wes} and in vibrating granular 
layers \cite{umb}. Although the physical
mechanism responsible for these patterns is different in each system
they can be described within a common framework. A hexagonal pattern can be seen as the 
superposition of three systems of rolls at $2 \pi/3$ rad, so that
the resonance condition ${\bf k_1}+{\bf k_2}+{\bf k_3} = 0$ is satisfied. 
The main aim of this
paper is to discuss the form of the evolution equations of the amplitudes 
of those three modes, 
ie., the so-called  {\em amplitude equations}, as well as the most general 
linear {\em phase equation} for a hexagonal pattern and the corresponding 
stability regions.

	From symmetry arguments one can deduce \cite{gol} that the
normal form for the amplitude of the modes forming the hexagonal pattern
is
\begin{equation}
\partial_t A_1 = \epsilon A_1 + \alpha_0\overline{A}_2\overline{A}_3
-\gamma(|A_2|^2+|A_3|^2)A_1 - |A_1|^2 A_1 
\label{eq:amp}
\end{equation}
(The overbar denotes complex conjugation. The equations for $A_2$ and 
$A_3$ are obtained by rotating the subindexes.)
Spatial variations can be included following the
Newell-Whitehead technique (see ref. \cite{new}). Up to the third order
in the amplitudes the linear spatial variations of each system of rolls
is in the form $({\bf\hat n_1}\cdot \nabla)^2 A_1$, a term that must be 
added to eq. (\ref{eq:amp}). Here ${\bf \hat n}_1$ indicates a unitary
vector in the direction of the first system of rolls.
Until recently, as in the case
of a pattern of rolls, only this term was considered. Brand \cite{bra}
discussed the possibility of including terms in the form $(A \nabla A)$. 
Considering the hexagonal symmetries (reflections on X-axis and Y-axis, and
$2 \pi/3$ rad rotations) this author deduced that a term in 
the form 
\begin{equation}
i \beta_1 [ \overline A_3 ({\bf \hat n}_2 \cdot \nabla) \overline A_2
+ \overline A_2 ({\bf \hat n}_3 \cdot \nabla) \overline A_3]
\label{eq:beta1}
\end{equation}
($\beta_1$ is a real coefficient) could be also added to eq. (\ref{eq:amp}).
However, as noticed by several authors \cite{gun}, \cite{kuz}, in the same order
another term can appear, namely
\begin{equation}
i \beta_2 [ \overline A_3 ({\bf \hat n}_3 \cdot \nabla) \overline A_2
+ \overline A_2 ({\bf \hat n}_2 \cdot \nabla) \overline A_3]
\label{eq:beta2}
\end{equation}
(again $\beta_2$
 is a real coefficient). It can easily be seen that 
this term also remains invariant under the hexagonal group 
transformations \cite{nota}. 

Then, for perturbations up to the third order a generalized 
amplitude equation 
that accounts for spatial variations in a hexagonal pattern is 
\begin{eqnarray}
\partial_t A_1 & = & \epsilon A_1 + 
({\bf\hat n_1}\cdot \nabla)^2 A_1  
 + \alpha_0\overline{A}_2\overline{A}_3+ \nonumber  \\
&+&i \beta_1 [ \overline A_3 ({\bf \hat n}_2 \cdot \nabla) \overline A_2
+ \overline A_2 ({\bf \hat n}_3 \cdot \nabla) \overline A_3]
+i \beta_2 [ \overline A_3 ({\bf \hat n}_3 \cdot \nabla) \overline A_2
+ \overline A_2 ({\bf \hat n}_2 \cdot \nabla) \overline A_3] - \nonumber \\
 &-&\gamma(|A_2|^2+|A_3|^2)A_1 -|A_1|^2 A_1
\label{eq:amp2}
\end{eqnarray}
\begin{figure}
\vbox to 4.5cm{\vfill\centerline{\fbox{Here is the figure 1. and 2.}}\vfill}
\caption{Unitary vectors: ${\bf n}_i$ parallel
and ${\bf \tau}_i$ perpendicular to the wavenumbers in a hexagonal pattern.}
\label{fig1}
\end{figure}
\begin{figure}
\caption{a) Dilatations and  b) distortions of a hexagonal pattern.}
\label{fig2}
\end{figure}
The gradient terms in eqs. (\ref{eq:beta1}) and (\ref{eq:beta2}) are a consequence 
of quadratic resonances and therefore their influence is esentially 
two-dimensional. (It should be noticed that, at the same order, terms in the
form $|A|^2\nabla A$ could be included, but for the shake of simplicity we will 
not consider them in this paper). To gain some 
physical insight into the problem it is useful to express the derivatives
in eq. (\ref{eq:beta2}) in terms of the unitary vectors of
the corresponding mode, i.e, $ {\bf \hat n}_2 = -\frac{1}{2} {\bf \hat n}_3
+ \frac{\sqrt 3}{2}{\bf \hat \tau}_3$ in the first term and 
$ {\bf \hat n}_3 = -\frac{1}{2} {\bf \hat n}_2 - 
\frac{\sqrt 3}{2}{\bf \hat \tau}_2$ in the second, where ${\bf \hat \tau}_i$
stands for the unitary vectors perpendicular to the direction of the 
wavenumber of the corresponding system of rolls (See fig. 1). Notice
that the gradient terms (eqs. (\ref{eq:beta1}) and (\ref{eq:beta2})) can 
be added to give 
\begin{equation}
i \alpha_1 [ \overline A_3 ({\bf \hat n}_2 \cdot \nabla) \overline A_2
+ \overline A_2 ({\bf \hat n}_3 \cdot \nabla) \overline A_3] +
i \alpha_2 [ \overline A_2 ({\bf \hat \tau}_3 \cdot \nabla) \overline A_3
-\overline A_3 ({\bf \hat \tau}_2 \cdot \nabla) \overline A_2]
\label{eq:tau}
\end{equation}
with $ \alpha_1 = \beta_1 - \frac{1}{2} \beta_2$ and $\alpha_2 = 
\frac{\sqrt 3}{2} \beta_2$. Terms in this form have been discussed by 
Gunaratne et al. \cite{gun} for chemical reactions, 
Kuznetsov et al. \cite{kuz} for Rayleigh-B\'enard convection and Bragard 
and Golovin et al. \cite{brag} for B\'enard-Marangoni convection. A term 
with the coefficient $\alpha_1$ accounts for distortions
in the direction of the rolls and it therefore corresponds to 
{\em dilatations}
of hexagons (it slightly changes the volume in Fourier space), while
the terms with $\alpha_2$ account for {\em distortions} of the hexagonal
form. In fig. 2 we represent the action of these two terms 
in Fourier space.
As discussed by Kuznetsov et al. \cite{kuz},  $\alpha_0$ and $\alpha_1$ 
vanish when the bifurcation that leads to the hexagonal pattern is 
supercritical, while in
the subcritical case the three coefficients $\alpha$ are, in general, different 
from zero. The term with $\alpha_1$  will stabilize patterns with
$|k| \neq |k_c|$ while the term with $\alpha_2$ would stabilize
non-equilateral hexagons. 

With the transformation in eq. (\ref{eq:tau}), eq. (\ref{eq:amp2})
becomes
\begin{eqnarray}
\partial_t A_1 & = & \epsilon A_1 + ({\bf\hat n}_1\cdot \nabla)^2 A_1  
 + \alpha_0\overline{A}_2\overline{A}_3 + \nonumber  \\
&+&i \alpha_1 [ \overline A_3 ({\bf \hat n}_2 \cdot \nabla) \overline A_2
+ \overline A_2 ({\bf \hat n}_3 \cdot \nabla) \overline A_3]
+i \alpha_2 [ \overline A_2 ({\bf \hat \tau}_3 \cdot \nabla) \overline A_3
- \overline A_3 ({\bf \hat \tau}_2 \cdot \nabla) \overline A_2] - \nonumber \\
 & -&\gamma(|A_2|^2+|A_3|^2)A_1 -|A_1|^2 A_1
\label{eq:amp3}
\end{eqnarray}
Now let us consider solutions with a wavenumber $k$
slightly different from $k_c$, i.e., $A_i = \hat A_i 
e^{i{\bf q}_i} \cdot {\bf r}$ with ${\bf q}_i ={\bf k}_i  - {\bf k}_c$. We are 
interested in homogeneous and stationary solutions of
the last equation in the form of hexagons $\hat A_1 = \hat A_2 = \hat A_3 = H 
\neq 0$ for which the amplitude must be 
\begin{equation} 
H = \frac{(\alpha_0 + 2 q \alpha_1) + \sqrt{(\alpha_0 + 2 q \alpha_1)^2
+ 4(\epsilon - q^2)(1 + 2\gamma)}}{2(1 + 2 \gamma)} 
\end{equation}
The stability of this solution is determined considering perturbations in the 
form $A_i = H e^{iq_i\cdot x_i}(1 + r_i + i\phi_i)$, where $r_i$ is the
amplitude and $\phi_i$ is the phase of the perturbation. After introducing
these perturbations in eq. (\ref{eq:amp3}) and linearizing one arrives at
the following system of equations
\begin{eqnarray}
\partial_t r_1 & =& \partial_1^2 r_1 - 2 q \partial_1 \phi_1 
+ (\alpha_0 + 2q \alpha_1) H (r_2 + r_3 - r_1) + 
H (\alpha_1 + \frac{\alpha_2}{\sqrt 3})
(\partial_2 \phi_2 + \partial_3 \phi_3) +\nonumber \\
&  &+ \alpha_2 H (\partial_3 \phi_2 
+ \partial_3 \phi_3) - 2H^2 r_1 - 2 \gamma H^2 (r_2+ r_3)\\
\partial_t \phi_1 & =& 2 q \partial_1 r_1 + \partial_1^2 \phi_1 
-(\alpha_0 + 2q \alpha_1) H (\phi_1 + \phi_2 + \phi_3) +
H (\alpha_1 + \frac{\alpha_2}{\sqrt 3})(\partial_2 r_2 + \partial_3 r_3)+  
\nonumber \\
&  &+ \frac{2}{\sqrt 3} \alpha_2 H (\partial_2 r_3 + \partial_3 r_2) 
\end{eqnarray}
where we have used the following notation $\partial_i = {\bf \hat n}_i 
\cdot \nabla$. We assume that the amplitudes $r_i$ and the total
phase $\Phi = \phi_1 + \phi_2 + \phi_3$ are fastly decaying variables
and therefore they can be eliminated adiabatically. As a result
the dynamics is dominated by two of the phases. Instead of using
$ \phi_2$ and $\phi_3$ one can take a vector ${\vec{\phi}} = [-(\phi_2 +\phi_3),
\frac{1}{\sqrt 3}( \phi_2 - \phi_3)]$ and the resulting equation will have the
most general form of a linear diffusion equation in 2D
\begin{equation}
\partial_t \vec {\phi} = D_{\bot} \nabla^2 \vec {\phi} + 
(D_{\|} -D_{\bot}) \nabla(\nabla \cdot \vec {\phi})
\end{equation}
This is the linear {\em phase equation} of a pattern of hexagons.
The form of the coefficients in this equation have been chosen by 
analogy with that in the wave equation in an elastic solid \cite{lan}. 
(The velocity
of the transversal waves $c_t$ corresponds here to $D_{\bot}$, while the  
velocity for the longitudinal waves $c_l$ is replaced by $D_{\|}$.)
Using this analogy we split the phase $\vec{\phi}$ into a longitudinal 
part $\vec {\phi}_l$ and a transversal $ \vec {\phi}_t$, that  
satisfy $\nabla \times \vec {\phi}_l = 0$  (rhomboidal phase perturbations) 
and $\nabla \cdot \vec{\phi}_t=0$ (rectangular phase perturbations).  
It can be proved straightforwardly that these components satisfy
\begin{equation}
\partial_t  \vec{\phi}_l = D_{\|} \nabla^2 \vec{\phi}_l, \;\;\;\;\;\;  
\partial_t  \vec{\phi}_t = D_{\bot} \nabla^2  \vec{\phi}_t
\end{equation} 
A linear stability analysis of these equations shows that the system
is stable to phase perturbations provided that $D_{\bot} > 0$ 
and $D_{\|} >0$. After tedious calculations one arrives to a general
expression of these coefficients
\begin{eqnarray}
D_{\bot} &=& \frac{1}{4} - \frac{q^2}{2u} + 
\frac{H^2}{8u} (\alpha_1 - {\sqrt 3}\alpha_2)^2\\ \label{eq:deper}
D_{\|} &=& \frac{3}{4} - \frac{q^2(4u+v)}{2uv} + 
\frac{H^2}{8u} (\alpha_1 - {\sqrt 3}\alpha_2)^2 - 
\frac{H^2 \alpha_1}{v}(\alpha_1 + {\sqrt 3} \alpha_2) + 
\frac{Hq}{v}(3 \alpha_1 + {\sqrt 3} \alpha_2)
\end{eqnarray}
with the relationships
\begin{eqnarray}  
&& u = H^2(1 - \gamma)+ (\alpha_0 + 2\alpha_1q)H > 0\\
&& v = 2H^2(1 + 2\gamma) - (\alpha_0 + 2\alpha_1q)H >0 \label{eq:v}
\end{eqnarray}
\begin{figure}
\vbox to 6.5cm{\vfill\centerline{\fbox{Here is the figure 3.}}\vfill}
\caption{Stability region of hexagonal cells with 
$\alpha_1 = \alpha_2 = 0$. Notice that this region is symmetrical 
respect to the vertical axis. The phase instabilities are represented 
by a solid ($D_|=0$) and a dashed line ($D_\bot=0$).}
\label{fig3}
\end{figure}
	The curves $D_{\bot} = 0$ and $D_{\|} =0$ determine the stability
of a regular hexagonal pattern to  rhomboid and rectangular phase 
perturbations respectively, while $u =0$ 
determines the region where the hexagons are unstable to amplitude
perturbations. Let us mention that we have assumed that the homogeneous
stationary solution $H$ is positive, so $ \alpha_0 + 2q\alpha_1 >
0$. Otherwise the hexagons become unstable to a global phase 
change from $0$ to $\pi$ (up-hexagons to down-hexagons). It is
interesting to examine the influence of the different parameters in 
phase stability. We first consider the particular case
in which the nonlinear spatial terms are absent ($\alpha_1 =\alpha_2 = 0$).
This case has been considered by several authors \cite{lau,hoy,sus}.
The results are given in fig. 3 in a $(q, \epsilon)$ representation, 
where the shaded area corresponds to the stability region of the
hexagonal pattern. 
Notice that due to the symmetry $q \rightarrow - q$ in the diffusion 
coefficients and in $u$ and $v$ this figure is symmetrical with respect to 
the vertical axis. 
Near threshold rhomboidal phase perturbations destabilize the pattern 
($D_{\|} = 0$), but for higher supercritical conditions
the pattern becomes unstable by rectangular phase perturbations ($D_{\bot} = 0$). 
These two curves intersect at the values of $q$ that correspond to the
conditions $D_{\bot} = D_{\|} = 0$, i.e., $q = \pm \frac{\alpha_0}{2 \gamma} 
\sqrt{(1 + \gamma)/2}$. (For $(\alpha_1, \alpha_2) \neq 0$ these
points are quantitatively but not qualitatively modified.) 
Both curves are tangent at $q = 0$ with the upper amplitude stability 
curve $u =0$

From eqs. (\ref{eq:deper})-(\ref{eq:v}) one can deduce the following 
symmetry properties
\begin{eqnarray}
H (\alpha_1, q) = H (- \alpha_1, - q) ;\;\;\ 
u (\alpha_1, q) = u (- \alpha_1, - q) ;\;\;\ 
v (\alpha_1, q) = v (- \alpha_1, - q)  \\ \nonumber
D_{\|} (\alpha_1, \alpha_2, q) = D_{\|} ( - \alpha_1, - \alpha_2, - q) ;\;\;\ 
D_{\bot} (\alpha_1, \alpha_2, q) = D_{\bot} ( - \alpha_1, - \alpha_2, - q)
\end{eqnarray}
These symmetry expressions imply that the stability curves for a particular 
value of $(\alpha_1, \alpha_2)$ become reflected in a $(q, \epsilon)$ 
representation with 
respect to the vertical axis under the transformation $\alpha_1 \rightarrow 
-\alpha_1$, $\alpha_2 \rightarrow -\alpha_2$. 
\begin{figure}
\vbox to 6.5cm{\vfill\centerline{\fbox{Here is the figure 4.}}\vfill}
\caption{Stability region of a hexagonal pattern when all
spatial terms are included. a) Closed case ($\alpha_0=1,\alpha_1=\alpha_2=0.5,
\gamma=4$). b) Open case ($\alpha_0=1,\alpha_1=1,\alpha_2=2.5,\gamma=4$).}
\label{fig4}
\end{figure}
The stability curves are qualitatively modified when the gradient
terms are present. In general, for 
$(\alpha_1, \alpha_2) \neq 0$ the stability regions are
no longer symmetrical with respect to the $\epsilon$-axis.  An example of this 
situation is given in fig. 4a.
Several cases are possible. We see that for $(\alpha_1, \alpha_2) > 0$ 
the phase instability curves ($D_{\|} = 0, D_{\bot} = 0)$ 
are decentered to the right, while the 
minimum of amplitude instability curve  $u = 0$ is decentered to the 
left in the $(q, \epsilon)$ 
plane. The phase instability and the amplitude instability
curves are tangent at $q = 0$ when the two conditions $D_{\bot} = 0$ and
$u = 0$ are met simultaneously. This leads to the condition $\alpha_1
= \sqrt {3}\alpha_2$. (Notice that the two phase instability curves are
tangent at the same point). But in many cases the phase and amplitude
instability curves intersect at two points, namely
\begin{equation}
q = - \frac{\alpha_0 (\alpha_1 - \sqrt{3} \alpha_2)}{ 2[\alpha_1
(\alpha_1 - \sqrt{3} \alpha_2))  \pm (\gamma - 1)]}
\end{equation}
However, when the following condition 
$|\alpha_1 (\alpha_1 - \sqrt{3} \alpha_2)| \geq (\gamma - 1)$ is satisfied 
the curve
$D_{\bot} = 0$ is not closed and the phase and amplitude instability
curves do not intersect themselves in the $q \geq 0$ quadrant. (Such
a situation is represented in fig. 4b)

When convection is due to surface-tension variations with the temperature 
(Benard-Marangoni (BM) convection) the condition $\alpha_1 > 0$ is satisfied 
\cite{blas}. Two examples of the the phase stability region 
for $(\alpha_1, \alpha_2) \geq 0$ are given in Fig. 3. 
We notice that the stability regions in Figs. 3a and 3b
are in qualitative agreement with the numerical results obtained
by Bestehorn \cite{bes} for the BM problem. 
We shall mention
that in experiments performed by Koschmieder \cite{kosch-swif} the number of cells
increases ($q$ increases) with the supercritical heating in BM convection. 
The corresponding
values are fitted quite well with the line of maximal growth rate 
obtained numerically by Bestehorn \cite{bes}. (This line remains inside the
phase stability region.)  This asymmetry could explain why in that system 
the transition between hexagons and rolls is not observed near threshold.
	
	From the general form of the amplitude equations 
we derived the {\em phase equation} for a hexagonal pattern 
which is formally similar to the
wave equation in an elastic solid. The
expressions of the coefficients in this equation allow to determine 
the stability diagram. Unfortunately, experimental results on these phase
instabilities are not yet available. We hope that the present
results will suggest new experiments to study the wavenumber selection
mechanisms in hexagonal patterns in different physical systems.

\stars

	Fruitful discussions with R. Hoyle (Cambridge), M. Bestehorn
(Stuttgart, Cottbus), A.G Golovin and A.E. Nuz (Haifa) 
are gratefully acknowledged. This work was partially financed 
by the DGICYT (Spanish Government) under grants PB95-0578
and by the PIUNA (Universidad de Navarra). One of us (B.E.) aknowleges 
the Basque Government for a fellowship (BFI95.035).


\vskip-12pt
\begin{thebibliography}{99}

\bibitem{ben}        
\Name{B\'{e}nard H.} \Review{Rev. Gen. Sci. Pures Appli.} \Vol{11} \Year{1900}
\Page{1261}. 

\bibitem{wol} 
\Name{Wollkind D. J., Sriranganathan R. \And Oulton D. B.} \Review{Physica D} 
\Vol{12} \Year{1984} \Page{215}.

\bibitem{cil-bod-ste} 
\Name{Ciliberto S., Pampaloni E. \And P\'erez-Garc\'{\i}a C.} \Review{Phys. Rev. Lett.}
\Vol{61} \Year{1988} \Page{1198}.  

\bibitem{fau-gol} 
\Name{Edwards W. S. \And S. Fauve} \Review{J. Fluid Mech.} \Vol{278} \Year{1994} 
\Page{123}.

\bibitem{d'al} 
\Name{D'Alessandro G. \And Firth W. J.} \Review{Phys. Rev. Lett.} \Vol{66} 
\Year{1992} \Page{2597}. 

\bibitem{pam} 
\Name{Pampaloni E., Ramazza P. L., Residori S. \And Arecchi F.T.} \Review{Europhys. Lett.}
\Vol{24} \Year{1993} \Page{587}.

\bibitem{dek-swin}   
\Name{de Kepper P., Castets V., Dulos E. \And Boissonade J.} \Review{Physica D} 
\Vol{49} \Year{1991} \Page{161}; \Name{Ouyang Q. \And Swinney H. L.} \Review{Chaos}
\Vol{1} \Year{1991} \Page{411}.

\bibitem{gai-wes}   
\Name{Gailitis A.} \Review{J. Fluid Mech.} \Vol{82} \Year{1987} \Page{401}; 
\Name{Abou B. \And Wesfreid J. E.} 
in preparation.

\bibitem{umb}   
\Name{Melo F., Umbanhoward P. \And Swinney H. L.} \Review{Phys. Rev. Lett.} 
\Vol{95} \Year{1995} \Page{3838}.

\bibitem{gol}   
\Name{Golubitsky M., Swift J. W. \And Knobloch E.} \Review{Physica D} \Vol{10} 
\Year{1984} \Page{249}. 

\bibitem{new}   
\Name{Newell A. C., Passot T. \And Lega J.} \Review{Ann. Rev. Fluid Mech.} 
\Vol{25} \Year{1993} \Page{399}.

\bibitem{bra}   
\Name{Brand H.} \Review{Progr. Theor. Phys., Suppl.}  \Vol{99} \Year{1989} \Page{442}.

\bibitem{gun}   
\Name{Gunaratne G., Ouyang Q. \And Swinney H. L.} \Review{Phys. Rev. E} \Vol{50} 
\Year{1994} \Page{2802}.

\bibitem{kuz}   
\Name{Kuznetsov E. A., Nepomnyashchy A. A. \And Pismen L. M.} \Review{Phys. Lett. A}
\Vol{205} \Year{1995} \Page{261}. 

\bibitem{brag}
\Name{Bragard, J.} Ph.D. Thesis, Universidad Complutense de Madrid (1996). 
Unpublished; \Name{Golovin. A.A., Nepomnyashchy, A.A. \And Pismen, L.M.} \Review{J. Fluid Mech.}
\Vol{25} \Year{1997} \Page{317}.

\bibitem{nota}
As discussed in ref. \cite{gun} terms in the form $i({\overline A}_{i+1} \nabla^2 
{\overline A}_{i+1})/2q_c$ and $i(\nabla{\overline A}_{i+1})\cdot(\nabla{\overline A}_{i+2})/q_c$
must be added to eqs. (\ref{eq:beta1}) and (\ref{eq:beta2}) to render the resulting equation rotationally
invariant. As we restrict our analysis up to third order in the amplitude
we will not consider these kind of terms.

\bibitem{lan}
\Name{Landau L. D. \And Lifshitz E. M.} \Book{Teor\'{\i}a de la Elasticidad} 
Edit. Revert\'{e}, Barcelona \Year{1969}.

\bibitem{lau}  
\Name{Lauzeral J., Metens S. \And Walgraef D.} \Review{Europhys. Lett.} \Vol{24}  
\Year{1993} \Page{707}. 

\bibitem{hoy} 
\Name{Hoyle R.} \Review{Appl. Math. Lett.} \Vol{9} \Year{1995} \Page{81}.

\bibitem{sus}
\Name{Sushchik M. M. \And Tsimring L. S.} \Review{Physica D} \Vol{74} \Year{1994} 
\Page{90}.

\bibitem{blas}
\Name{Echebarria B. \And P\'erez-Garc\'{\i}a C.} in preparation.

\bibitem{bes}
\Name{Bestehorn M.} \Review{Phys. Rev. E} \Vol{48} \Year{1993} \Page{3622}.

\bibitem{kosch-swif}
\Name{Koschmieder L. \And Switzer D. W.} \Review{J. Fluid Mech.} \Vol{240} 
\Year{1992} \Page{533}.

\end{thebibliography}
\end{document}